\documentclass[10pt]{article}

\usepackage[
    letterpaper,
    textwidth=6.75in,
    textheight=9.0in,
    left=0.75in,
    top=1.0in,
    columnsep=0.25in
]{geometry}

\usepackage[T1]{fontenc}
\usepackage{times}
\usepackage{microtype}

\makeatletter
\renewcommand{\normalsize}{%
    \@setfontsize\normalsize{10pt}{12pt}%
    \abovedisplayskip 7pt plus 2pt minus 5pt
    \belowdisplayskip 7pt plus 2pt minus 5pt
    \abovedisplayshortskip 0pt plus 3pt
    \belowdisplayshortskip 4pt plus 3pt minus 3pt
}
\makeatother

\normalsize

\usepackage{titlesec}

\titleformat{\section}
    {\normalfont\fontsize{12}{14}\bfseries\scshape\raggedright}
    {\thesection}{1em}{}

\titlespacing*{\section}
    {0pt}{0.12in}{0.02in}

\titleformat{\subsection}
    {\normalfont\normalsize\bfseries\raggedright}
    {\thesubsection}{1em}{}

\titlespacing*{\subsection}
    {0pt}{0.10in}{0.01in}

\titleformat{\subsubsection}
    {\normalfont\normalsize\itshape\raggedright}
    {\thesubsubsection}{1em}{}

\titlespacing*{\subsubsection}
    {0pt}{0.08in}{0.01in}

\usepackage{graphicx}
\usepackage{subfigure}
\usepackage{booktabs}
\usepackage{float}
\usepackage[table]{xcolor}
\usepackage{caption}

\newcommand{\figCaptionSpacing}{1}

\usepackage{amsmath}
\usepackage{amssymb}

\usepackage{enumitem}

\setlist[itemize]{
    topsep=4pt plus 1pt minus 2pt,
    partopsep=1pt plus 0.5pt minus 0.5pt,
    itemsep=2pt plus 1pt minus 0.5pt,
    parsep=2pt plus 1pt minus 0.5pt
}

\setlist[enumerate]{
    topsep=4pt plus 1pt minus 2pt,
    partopsep=1pt plus 0.5pt minus 0.5pt,
    itemsep=2pt plus 1pt minus 0.5pt,
    parsep=2pt plus 1pt minus 0.5pt
}

\usepackage{natbib}

\setcitestyle{
    authoryear,
    round,
    citesep={;},
    aysep={,},
    yysep={;}
}

\renewcommand{\cite}[1]{\citep{#1}}

\usepackage{courier}
\usepackage[most]{tcolorbox}
\usepackage{xurl}
\usepackage{hyperref}

\hypersetup{
    pdftitle={CPU Micro-Architectural Characteristic Sensitivity Study on CPU-Side Agentic Task Latency},
    pdfauthor={Jack Adiletta, Shixin Ji, Zhuoping Yang, Peipei Zhou},
    colorlinks=true,
    linkcolor=blue,
    citecolor=blue,
    urlcolor=blue
}

\definecolor{Takeawaycolor}{HTML}{2563EB}

\newtcolorbox{takeaway}{
    enhanced,
    colback=Takeawaycolor!8,
    colframe=Takeawaycolor,
    boxrule=0pt,
    borderline west={3pt}{0pt}{Takeawaycolor},
    arc=4pt,
    outer arc=4pt,
    left=7pt,
    right=5pt,
    top=5pt,
    bottom=5pt,
    before upper={\textbf{Takeaway:}\ },
    fontupper=\small
}

\newcommand{\printkeywords}{
    \begingroup
    \renewcommand{\thefootnote}{}
    \footnotetext{
        \textbf{Keywords:}
        Agentic AI, CPU Performance,
        Edge Computing, Performance Characterization,
        Microarchitecture

    }
    \endgroup
}

\renewcommand{\footnoterule}{
    \kern -3pt
    \hrule width 0.8in
    \kern 2.6pt
}

\begin{document}


\twocolumn[

\begin{center}

{\fontsize{14}{16}\selectfont\bfseries
CPU Micro-Architectural Characteristic Sensitivity Study on CPU-Side Agentic Task Latency
\par}

\vspace{0.25in}

{\fontsize{10}{12}\selectfont
Jack Adiletta, Shixin Ji, Zhuoping Yang, and Peipei Zhou
\par}

\vspace{0.05in}

{\fontsize{10}{12}\selectfont
Brown University
\par}

\end{center}

\vspace{0.20in}


\noindent
\centerline{\large\bf\scshape Abstract}

\vspace{-0.12in}

\begin{quote}
{
\fontsize{10}{12}\selectfont
We perform a novel study of CPU micro-architectural characteristics on how they affect agentic AI CPU side latency. Small enterprises and individual users looking to reduce costs and seeking data privacy benefit from running agents locally, motivating an investigation into the optimal performance characteristics of an edge CPU to improve user experience. We investigate the relationship between CPU core frequency, CPU uncore frequency, active physical cores, and symmetric multi threading with agentic task latency in single and multi agent configurations. Across eight experiments on an Intel edge CPU, we find that P-core frequency produces the largest median latency improvement among the evaluated parameters. Increasing P-core frequency from 2.5 to 5.5 GHz yields a 1.154$\times$ speedup for a single agent and a 1.341$\times$ speedup for eight concurrent agents. We are the first to characterize hardware micro-architectural relationships on  single and multi agent configurations with TerminalBench; we also provide an open source framework and dataset for repeatable CPU side latency agentic benchmarking.

}
\end{quote}

\vspace{0.12in}

]


\printkeywords


\section{Introduction}


AI agents utilize tools and iterative reasoning to perform diverse and difficult tasks such as software engineering, information retrieval, and web search. Agentic AI creates a new CPU--GPU paradigm which is not well characterized by current LLM benchmarking techniques~\cite{lau2026agentperfbenchbenchmarkingevaluationsuite, yang2026architecturalimplicationsagenticai}. Unlike conventional LLM inference, agentic applications repeatedly alternate between GPU-based model inference and CPU-side orchestration and tool execution. Prior work has established the need for CPU side execution characterization~\cite{raj2026understandinganalyzingoptimizingagentic,yang2026architecturalimplicationsagenticai}; we investigate how CPU resources and micro architectural operating points influence agentic performance.

We conduct our Agentic Investigation with a high performance edge system, running agentic tools and orchestration on the CPU while hosting an LLM on a local GPU with vLLM. Local agents are advantageous for two reasons:
\begin{itemize}
\vspace{-3mm}
    \item Locally hosted agentic systems can process sensitive information without sending data to external LLM providers, improving data sovereignty. This is critical for systems operating with secure data, small and medium enterprises, or individuals seeking data privacy~\cite{knoop2026privatellminferenceconsumer}.
    \item There is existence proof that hosting an LLM locally is financially beneficial for a small enterprise with savings incurred in less than four months of deployment~\cite{knoop2026privatellminferenceconsumer}.
\vspace{-2mm}
\end{itemize}


\begin{table}[t]
\centering
\caption{Comparison of CPU architectural characteristics
investigated in prior agentic AI research.}
\label{tab:cpu_characterization_comparison}

\resizebox{\columnwidth}{!}{%
\setlength{\tabcolsep}{3pt}
\begin{tabular}{lcccccc}
\toprule
\textbf{CPU Characteristic}
& \textbf{Raj}
& \textbf{Yang}
& \textbf{Choi}
& \textbf{Chang}
& \textbf{Yuan}
& \cellcolor{green!15}\textbf{Ours} \\
\midrule

Physical Core Count
& \checkmark & \checkmark & $*$ & $\times$ & $\times$
& \cellcolor{green!15}\checkmark \\

Logical Core Allocation
& $*$ & $*$ & \checkmark & $\times$ & $\times$
& \cellcolor{green!15}\checkmark \\

SMT Sensitivity
& $*$ & $\times$ & $\times$ & $\times$ & $\times$
& \cellcolor{green!15}\checkmark \\

CPU Core Frequency
& $\times$ & $\times$ & $\times$ & $\times$ & $\times$
& \cellcolor{green!15}\checkmark \\

CPU Uncore Frequency
& $\times$ & $\times$ & $\times$ & $\times$ & $\times$
& \cellcolor{green!15}\checkmark \\


CPU Memory Bandwidth
& $*$ & $*$ & $\times$ & $*$ & $\times$
& \cellcolor{green!15}\checkmark \\

\bottomrule
\label{comparison}
\end{tabular}%
}

\vspace{-2mm}
\footnotesize
\textit{Note:} $*$ indicates partial characterization or
insufficient detail to determine if the
characteristic was experimentally investigated.
\vspace{-5mm}

\end{table}

We build our investigation on recent work investigating the emerging CPU--GPU relationship created by agentic AI. The work in~\cite{raj2026understandinganalyzingoptimizingagentic} approaches agentic serving from a systems perspective, focusing on CPU side bottlenecks; the work found Agentic AI tool execution consumes a significant chunk of end-to-end latency. Further, systems with high performance GPUs shift the bottleneck to the CPU side. Finally, parallelization strategies on the CPU side to boost CPU utilization, like increased processes, eventually saturates overall throughput, limiting the work supplied to the GPU and reducing utilization of costly GPU resources~\cite{raj2026understandinganalyzingoptimizingagentic}. The work in~\cite{yang2026architecturalimplicationsagenticai} amplifies these findings: conventionally provisioned servers are not well equipped handle the fragmented nature of agentic serving. This fragmentation causes CPUs and GPUs to remain underutilized while experiencing short bursts of near-saturation. Meanwhile, the host side roles including scheduling, orchestration, and tool use demonstrate distinct resource profiles.

The server class solution proposed and deployed in~\cite{yang2026architecturalimplicationsagenticai}, Agora, employs harvesting CPU resources with an additional harvesting application on top of the agentic serving. While this is a viable option for improving server CPU utilization, harvesting requires useful co-located work and primarily improves resource utilization rather than accelerating the agent's CPU-side execution.

Huang et al.~\cite{huang2026swarmxagenticschedulinglowlatency} propose SwarmX, an advanced scheduling framework evaluated on production infrastructure comprising nearly one million CPU cores and 1,000 GPUs. Neural predictors estimate resource demands and predict LLM call latency during agentic applications, enabling effective request scheduling and compute allocation. SwarmX significantly improved tail latency and throughput for a large-scale deployment, however, these benefits may be unattainable for locally hosted agentic systems operated by individuals or enterprises. Local systems may not have the large pools of heterogeneous compute resources necessary to employ a cluster-level optimization strategy like SwarmX. Furthermore, scheduling improvements do not directly address the underlying CPU performance characteristics that determine the latency of individual agentic execution phases.

Prior work has investigated agentic performance on parts with different core counts~\cite{raj2026understandinganalyzingoptimizingagentic}, while others have investigated fleet wide utilization of CPU cores~\cite{yang2026architecturalimplicationsagenticai}. Yuan et al. experimentally investigated the breakdown of tool execution, finding that agents initially read and explore, then shift to execution and writing behavior. What their work does not isolate is how the CPU resources themselves affect or are affected by the progression of the agentic workload~\cite{yuan2026agenticaiworkloadcharacteristics}. 

Choi et al. experimented with different core allocations on an Azure VM with 24 VCPUs on an AMD EPYC 7V13 part~\cite{choi2026aiagentsequalcharacterizing}. They report a 13x latency reduction for the CPU-sensitive HotpotQA workload when increasing allocation from 2 to 8 cores under one LLM-latency configuration. Although their experiments demonstrate substantial sensitivity to CPU resource allocation, their investigation does not establish the relationship between physical cores, simultaneous multi threading (SMT), and logical cores. Among the studies in Table~\ref{comparison}, their work provides the most direct investigation of incremental CPU resource provisioning on a server class part. 

Additional work in~\cite{chang2026llminferenceagenticworkloads} generates a new benchmark, AgentSysBench, and records CPU memory, disk, and network utilization, during benchmarking. They examine eight concurrent Codex execution sandboxes, comparing disjoint 16 core sets against a shared 128 core CPU resource allocation. They identify a 1.22x slowdown when the resources are shared, but this work does not isolate the underlying CPU architectural parameters responsible for the observed performance degradation~\cite{chang2026llminferenceagenticworkloads}. The authors of~\cite{chang2026llminferenceagenticworkloads} explicitly identify that attributing the slowdown incurred by shared resource contention to a specific hardware component remains an open question.

Collectively, these studies demonstrate that CPU resource availability and contention influence agentic performance, but do not systematically isolate the sensitivity of agentic workloads to CPU core count, SMT, core frequency, and uncore frequency. Our work seeks to fill this gap and provide 3 novel contributions to the study of agentic AI workloads, and an accompanying open source dataset and test framework.

\begin{enumerate}
\vspace{-5mm}
    \item Single Agent CPU side task latency on an unconstrained CPU is most sensitive to CPU core frequency, but mostly unaffected by bandwidth reduction or increased memory latency.
\vspace{-3mm}
    \item Multiple agents running concurrently on a system drain shared resources as more agents-core pairs are activated with a performance drop off beginning at four cores with four agents.
\vspace{-3mm}
    \item Task latency for multiple concurrent agents is sensitive to uncore frequency and more sensitive to core frequency than single agents.
\vspace{-3mm}
    \item We provide an open source dataset and framework of recorded Qwen3 trajectories for repeatable benchmarking. The dataset contains recorded trajectories for Terminal Bench, ADE Bench, DAB step, GAIA, and Replication Bench, covering a variety of agentic operation. Results presented in this work are exclusively from Terminal Bench for close inspection of coding related tool calling and agentic overhead.
\vspace{-5mm}
    
\end{enumerate}

\section{Experimental Setup}

We systematically isolate variables to perform our sensitivity study of CPU resources. We will assess different CPU resources effects on Agentic AI CPU side task latency. The CPU resources we will test with our experimental setup are: CPU performance cores (P-cores) count, utilizing SMT, active core frequency, and uncore frequency. We test two software configurations: deploying a single agent across a CPU resource allocation, and deploying agents in a 1:1 ratio with physical P-cores.

\subsection{Hardware}

We use an ASUS Pro WS W680-ACE IPMI motherboard with liquid cooling as the base for the test bed. The system contains 2x32 GB DDR5 operating at the JEDEC default 4800 MT/s.

\textbf{CPU:} Our CPU under test is the Intel Core i9-14900KS comprising eight P-cores and sixteen efficient cores (E-cores)~\cite{Intel_CPU}. Intel's two core types, P-cores and E-cores both implement the x86 ISA, however, the P-cores have two logical cores implemented via SMT while the E-cores are single threaded (ST) cores, placed in clusters of 4~\cite{Intel_CPU}. 
For our experimentation, we only focus on the P-cores; the E-cores remain enabled with a constant configuration across all experiments, but the E-cores are excluded from the benchmark container CPU affinity masks. The P-cores have an experimentally verified configurable all-core frequency range (800 MHz - 5.9 GHz), and a configurable uncore frequency range (800 MHz - 5 GHz). We analyze hardware performance counters to verify the requested frequency. When a workload launches, we reduce all inactive physical cores to minimal frequency, but they still can perform background OS work. Inactive logical cores with an active sibling will operate at their sibling frequency.

\begin{figure}[t]
    \centering
    \includegraphics[width=\linewidth]{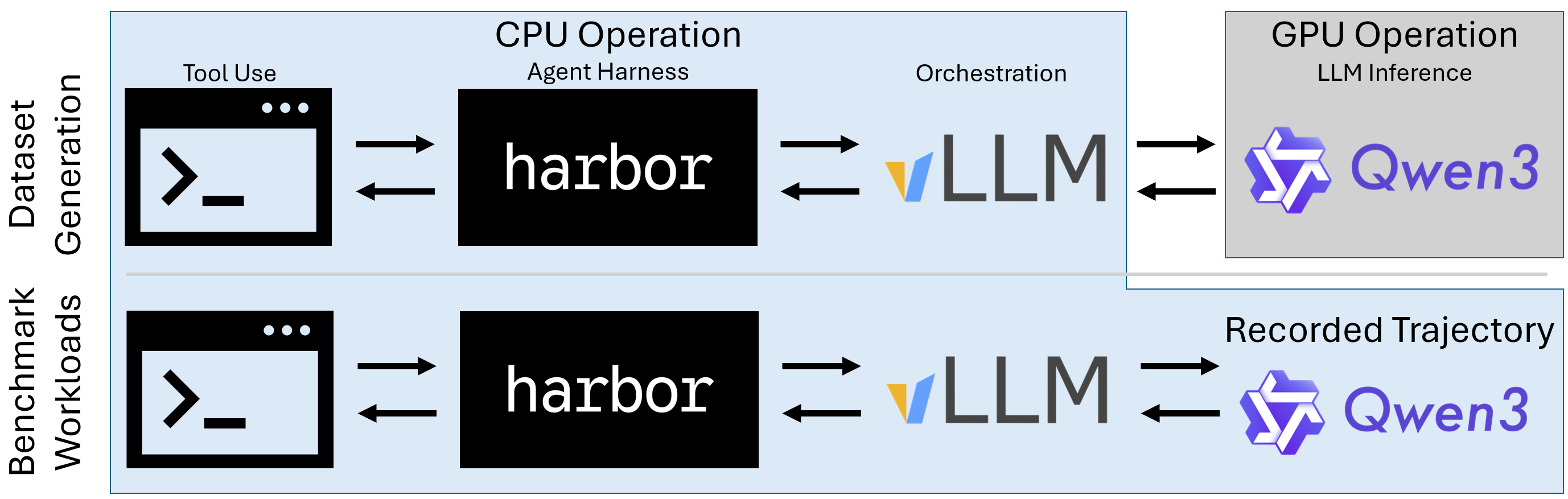}
    \vspace{-.75cm}
    \caption{NorthStar Replayable Benchmark}
    \vspace{-.6cm}
    \label{fig:ReplayableBench}
\end{figure}

\textbf{GPU:} The chosen GPU for building the Agentic workload dataset is the AMD Radeon PRO R9700 with 32 GB of GDDR6. We will not profile any local GPU operations as the focus of our study is CPU side operation.

\textbf{Exposing Experimental Variables:} We configure a specific P-core allocation by assigning Docker containers to the logical cores associated with P-cores. We configure SMT not by disabling the feature in BIOS, which would require repeated machine reboots, but by allocating 1 or 2 logical cores per physical core across the 1--8 physical cores. When utilizing SMT on P-core 0, we assign the container to logical cores 0 and 1 (we verified this mapping with Intel VTune). To prevent the workload from utilizing SMT, we assign the Docker container to logical core 0. To reduce additional work on the unutilized sibling logical core, we conduct no other work on the machine during benchmarking. This is less deliberate than disabling SMT through the BIOS, however, Intel VTune measurements report approximately 0\% utilization on excluded SMT siblings during benchmark execution. We request a fixed P-core frequency by setting the minimum and maximum core frequency limits exposed by the Linux CPU frequency interface to the same value -- we empirically verify this requested frequency prior to experimentation. We do not actively poll the cores during workload operation as that would incur overhead. We use a similar process for uncore frequency through the Linux \texttt{intel\_uncore\_frequency} driver.


\subsection{Software}

We have two requirements of our setup: each benchmark software component \textit{must} operate in a controlled hardware allocation, and the benchmarks must be repeatable. 

\textbf{Agentic Benchmark:} Our benchmark framework utilizes an unmodified Harbor Framework, a modified vLLM Docker container, and an additional Docker container to launch Harbor. Harbor is a popular agentic benchmarking tool~\cite{Harbor_Framework} used in prior agentic AI benchmarking work~\cite{yuan2026agenticaiworkloadcharacteristics}. A Docker compose file specifies the Harbor agentic sandbox configuration with the same set of controlled CPU resources. In this way, each benchmark component operates on a configurable CPU resource allocation.

\begin{figure}[t!]
    \centering
    \includegraphics[width=\linewidth]{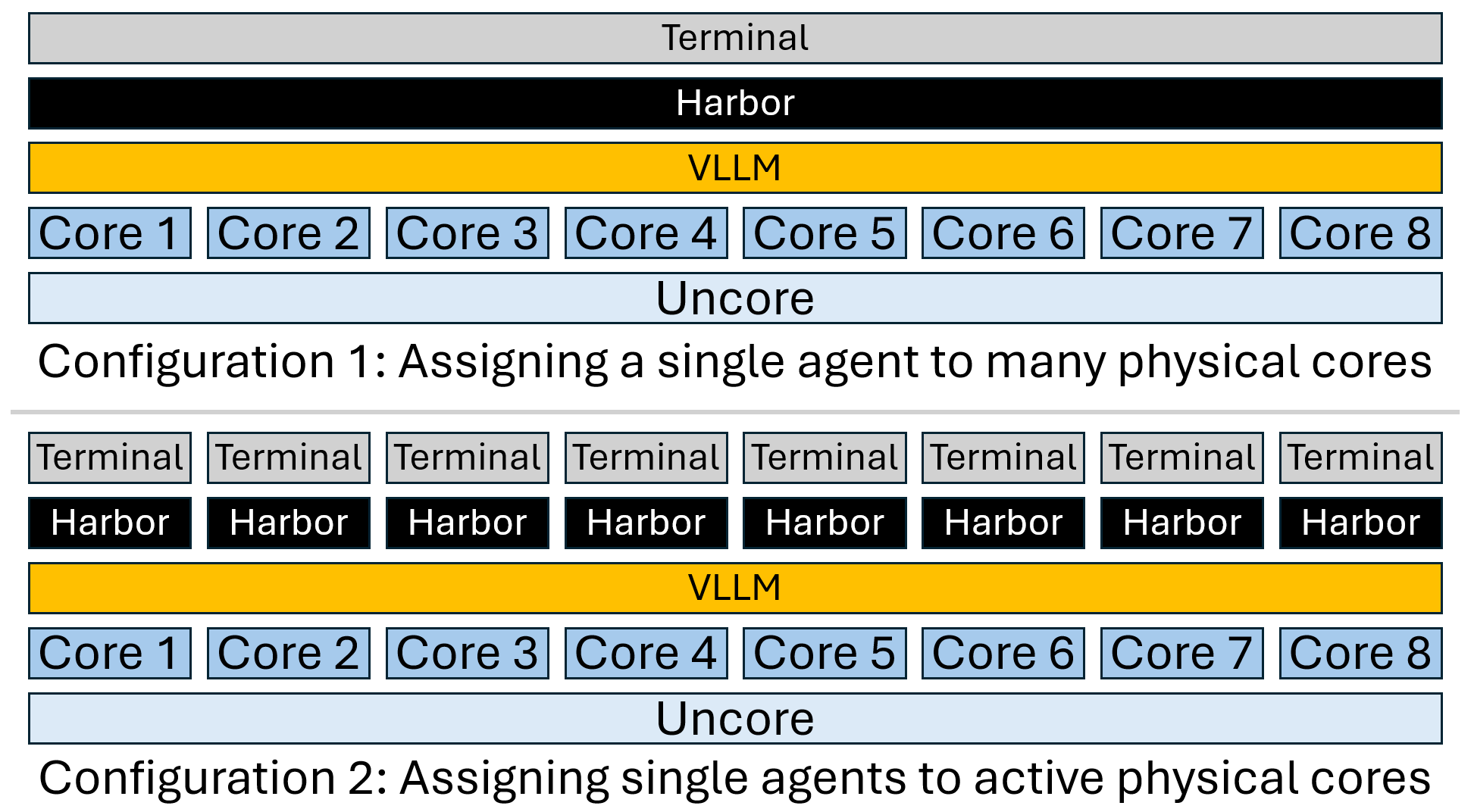}
    \vspace{-.75cm}
    \caption{Single Agent and Multi Agent Software Configurations}
    \vspace{-.6cm}
    \label{fig:software_config}
\end{figure}

\textbf{vLLM Serving:} vLLM is widely used in agentic workload characterization and serving-system evaluation~\cite{raj2026understandinganalyzingoptimizingagentic, yang2026architecturalimplicationsagenticai}. We choose to utilize a dark-side-of-the-code/Qwen3-Coder-30B-A3B-Instruct-AWQ model, a mixture-of-experts (MoE) model whose quantized weights fit within the GPU's 32 GB memory capacity with additional capacity for KV cache; prior agentic studies evaluated similarly sized Qwen3-family models~\cite{yuan2026agenticaiworkloadcharacteristics, wang2026benchmarkingllmservingsystems, lau2026agentperfbenchbenchmarkingevaluationsuite}. We conduct our experimentation in two phases: trajectory collection and deterministic replay, detailed in Figure~\ref{fig:ReplayableBench}. First we generate a set of agent trajectories for Terminal-Bench, an established agentic benchmark for characterizing agentic software development~\cite{yuan2026agenticaiworkloadcharacteristics, chang2026llminferenceagenticworkloads, lau2026agentperfbenchbenchmarkingevaluationsuite}. We record tokens and text produced by the model. We evaluate 39 of 47 Terminal-Bench tasks which completed their Terminal Bench task without an LLM caused error. During benchmarking we utilize a custom vLLM Docker image which injects the prerecorded responses in the vLLM datapath, circumventing LLM autoregressive generation. We verify that input tokens and injected model responses match that underlying dataset. Our modification intercepts vLLM requests immediately before invocation of the inference engine, preserving frontend request-processing, while replacing autoregressive generation with prerecorded model responses and modeled inference latency.

\textbf{A Replayable Benchmark:} Agentic workloads do not necessarily have deterministic operation~\cite{wang2026benchmarkingllmservingsystems}. By injecting prerecorded LLM responses, we force our benchmarks to use the identical tokenization inputs and tool calls. We approximate synthetic LLM decoding latency on a GPU representative of a local, high-performance GPU serving scenario~\cite{knoop2026privatellminferenceconsumer}. During benchmarking, we linearly interpolate measured time per output token (TPOT) at different concurrency levels based on a Hugging Face study of a similarly sized Qwen3-Omni-30B-A3B Instruct model on a high performance NVIDIA B200 reference system~\cite{jin2026qwen3omni}. Our modified vLLM tracks the number of in-flight vLLM requests, which we use as input to a linear interpolation of TPOT as a function of concurrent requests. The modified vLLM selects the TPOT at request initiation, fixed until request completion. This models increasing per-request inference latency under higher concurrency. We assign a fixed 50 ms synthetic GPU prefill delay, selected as an illustrative operating point informed by published TTFT measures rather than a direct measured GPU prefill duration~\cite{jin2026qwen3omni}, which reports TTFT but does not separately identify the contributions of CPU tokenization and GPU prefill. Prior work demonstrates that CPU tokenization can account for up to 49\% of TTFT for a prefix-cached Qwen3-32B workload~\cite{siavashi2026takingtokenizationhostagentic}. Because our replay mechanism preserves frontend tokenization, we retain this CPU-side contribution as actual tokenization execution time on the CPU while approximating GPU prefill through a fixed delay.

Our fixed prefill delay does not capture variations caused by prompt length, prefix-cache behavior, or GPU scheduling. TPOT for agent turns can change between benchmark runs as concurrency increases -- an artifact of the simulation environment intended to model higher TPOT as concurrency increases. The same in-flight tracking and TPOT policy is applied for every benchmark.

\textbf{Exposing Experimental Variables:} We configure the software in two ways, demonstrated in Figure~\ref{fig:software_config}. Configuration 1 deploys one agent (agent harness and terminal) to complete the benchmark, assigning the agent to the entire CPU allocation. Configuration 2 will launch one agent (agent harness and terminal) for each active physical core. In either configuration, a single vLLM session has access to all allocated CPU resources. Configuration 2 supported two sub configurations: shared resource allocation -- Harbor controls concurrency and the OS schedules all work among shared resources; and partitioned resource allocation -- a Docker container launches on each physical core, manually separating resources. Chang et al. observes a $1.22 \times$ slowdown when concurrent agent sandboxes share a CPU allocation compared to disjoint hardware configurations~\cite{chang2026llminferenceagenticworkloads}. This motivates testing these concurrency configurations on a smaller CPU platform.


\newpage

\begin{figure}[t]
    \centering
    \includegraphics[width=1\linewidth]{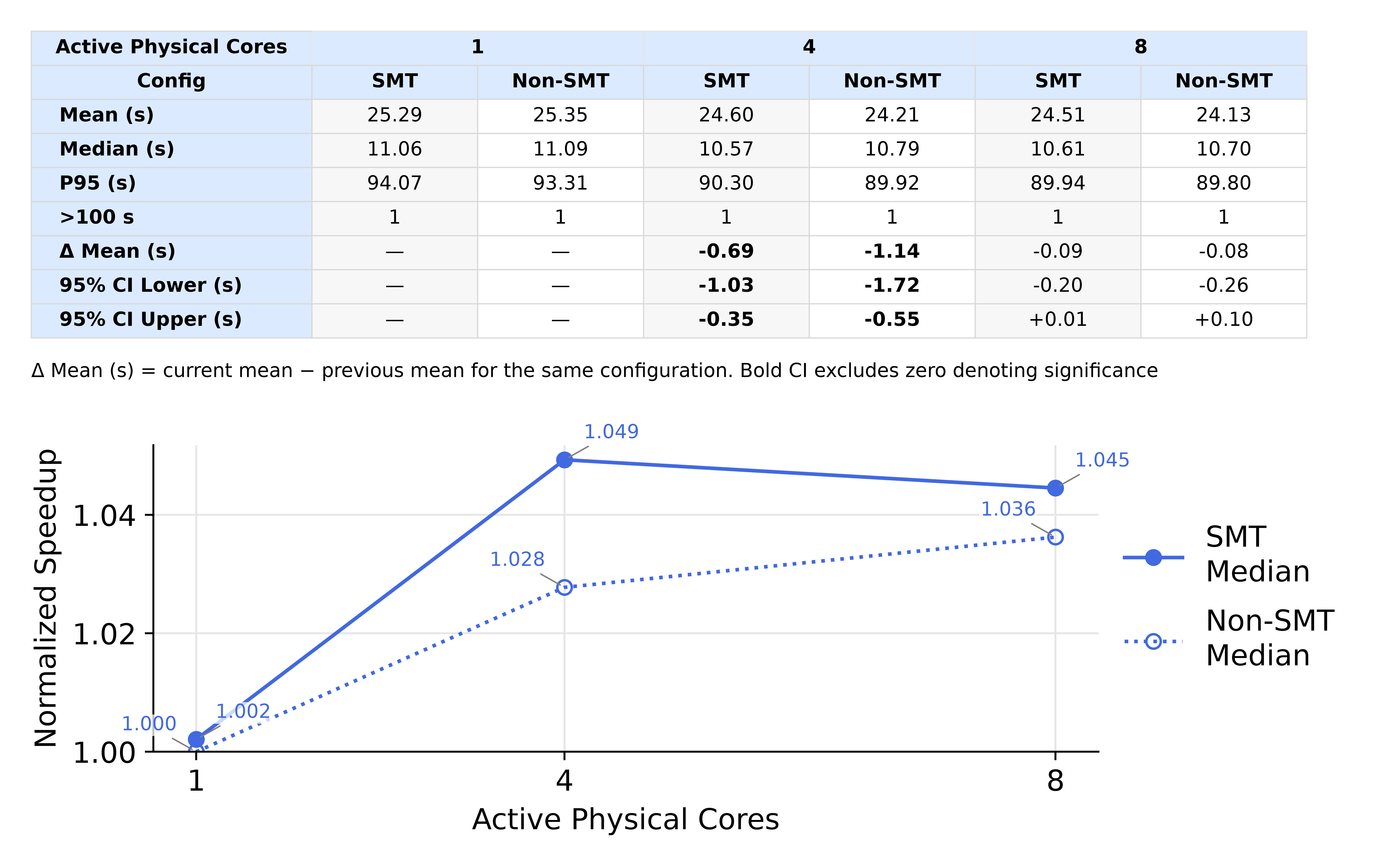}
    \vspace{-\figCaptionSpacing em}
    \caption{Single agent on increasing core allocations.}
    \vspace{-\figCaptionSpacing em}
    \label{fig:single_agent_core_sweep}
\end{figure}

\textbf{Experiment 1: Sweeping physical cores allocated to a single agent.} 
In this experiment, we launch one agent within a specified CPU allocation. The CPU allocations under test are 1, 4, and 8 P-cores while holding the core frequency at 4.5 GHz and the uncore frequency at 3 GHz. We test each configuration with and without SMT utilization.

Our experiments assess CPU side latency of a single agent software stack. We collect the latency of each agent task trajectory and find a total duration for each task -- this duration includes all CPU side activity and synthetic GPU latency. We remove GPU latency from total latency by subtracting the recorded delays for each agentic task.

The upper table reports mean, median, and 95th-percentile CPU-side task latency, and the number of tasks exceeding 100 seconds. We report the mean paired task-latency difference between adjacent P-core allocations within each SMT configuration, along with its 95\% confidence interval calculated using a paired t-test. Negative differences indicate reduced latency, while bold values indicate statistically significant differences ($p<0.05$). In Figure~\ref{fig:single_agent_core_sweep}, the lowest median latency is 10.57 seconds per task under the four P-core SMT configuration, a 0.49-second reduction compared with the single-core configuration.

The median task length speedup, relative to the single core, non-SMT median latency, for the three core allocations is shown in the lower plot of Figure~\ref{fig:single_agent_core_sweep}. Four P-cores with SMT has the greatest speedup relative to the single core non-SMT median task latency while speedup decreases slightly from four cores to eight cores with SMT. This could indicate diminishing returns beyond four physical cores for a single agent, even with SMT siblings available.

\begin{figure}[!b]
\vspace{-5mm}
    \begin{takeaway}
        Increasing P-core allocation from 1 to 4 cores generates a 1.049x and 1.028x speedup for SMT and non-SMT respectively. The speedup from 4 to 8 cores is smaller for non-SMT and decreases with SMT utilization indicating minimal gains from increased allocated physical resources beyond four cores for a single agent.
    \end{takeaway}
    \vspace{-5mm}
\end{figure}

\newpage

\begin{figure}[t]
    \centering
    \includegraphics[width=1\linewidth]{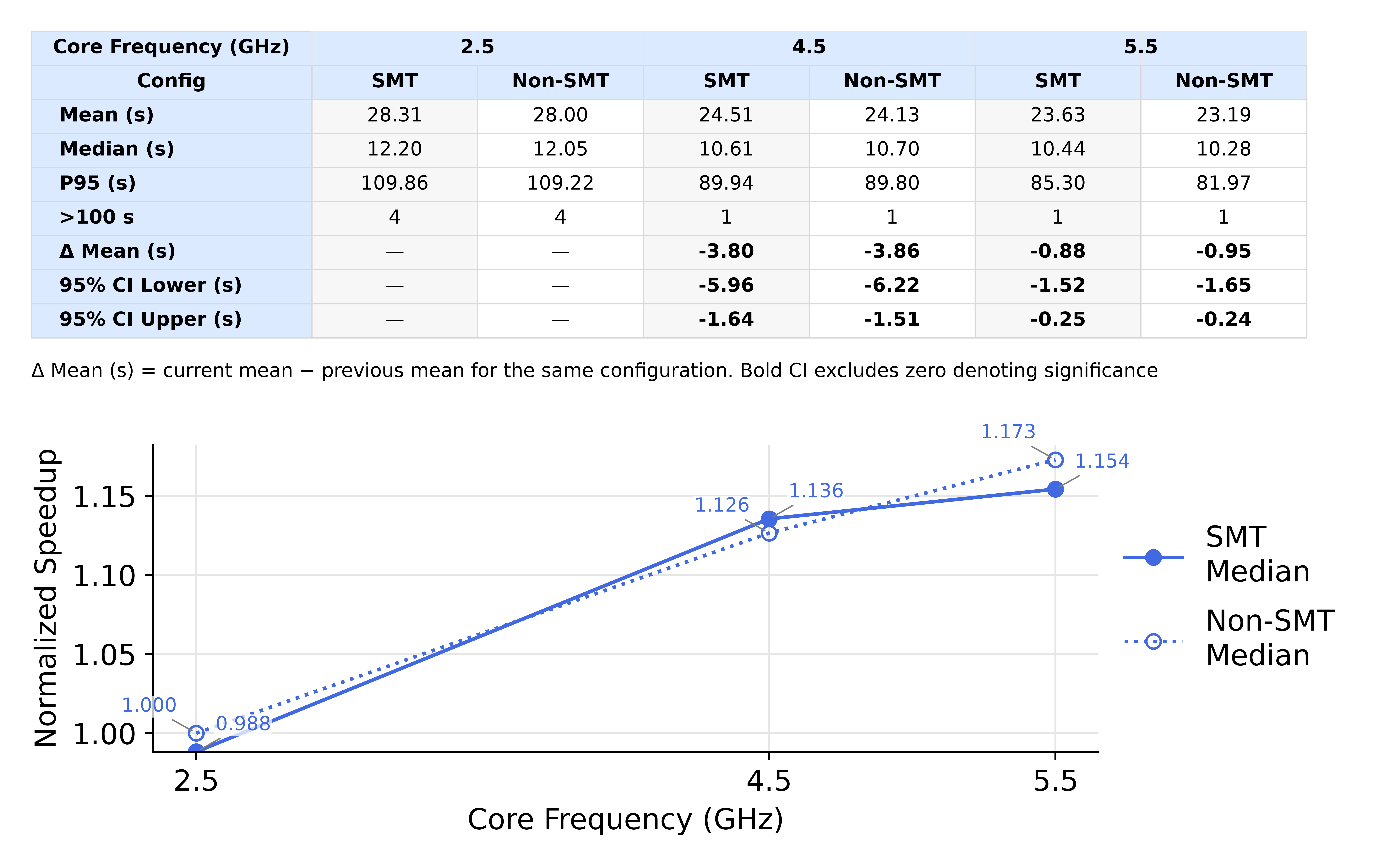}
    \vspace{-\figCaptionSpacing em}
    \caption{Single agent with increasing core frequency.}
    \vspace{-\figCaptionSpacing em}
    \label{fig:single_agent_core_freq_sweep}
\end{figure}

\textbf{Experiment 2: Single-agent P-core frequency sensitivity.} 
This experiment allocates all eight available P-cores to the agentic software stack. We vary P-core frequency with cores set to 2.5 GHz, 4.5 GHz, and 5.5 GHz. The uncore frequency is held at 3 GHz. The 4.5 GHz configuration corresponds to the eight P-core configuration evaluated in Experiment 1.

The upper table of Figure~\ref{fig:single_agent_core_freq_sweep} highlights the long-tailed nature of agentic execution. The number of tasks exceeding 100 seconds of CPU-side latency decreases from four at 2.5 GHz to one at both 4.5 GHz and 5.5 GHz.

The mean task latencies decrease from over 28.31 and 28.00 seconds for SMT and non-SMT respectively at a 2.5 GHz core frequency to 23.62 and 23.19 seconds at 5.5 GHz. The approximately five second decrease in average task latency from 2.5 GHz to 5.5 GHz compared with the roughly two second decrease in median task latency indicates increased core frequency reduced tail latencies more substantially than overall latencies.

Still, there is a 1.154$\times$ median latency speedup from 2.5 GHz to 5.5 GHz with non-SMT relative to the 2.5 GHz non-SMT median task latency. This speedup is brought on by a 120\% increase in core frequency, a limited latency sensitivity compared to core frequency. 

Increasing core allocations exhibit diminishing returns in experiment 1; limited frequency sensitivity of agentic workloads suggests that execution time is not exclusively determined by core compute performance. We turn our investigation to the memory hierarchy of our workstation, starting with the CPU uncore frequency domain to assess whether shared cache and interconnect operating points contribute to agentic task latency.

\begin{figure}[!b]
\vspace{-5mm}
    \begin{takeaway}
        Increasing P-core frequency by 120\% produces a median latency speedup of 1.154$\times$ for a single agent using eight P-cores, indicating limited frequency scaling.
    \end{takeaway}
\vspace{-5mm}
\end{figure}

\newpage

\begin{figure}[t]
    \centering
    \includegraphics[width=1\linewidth]{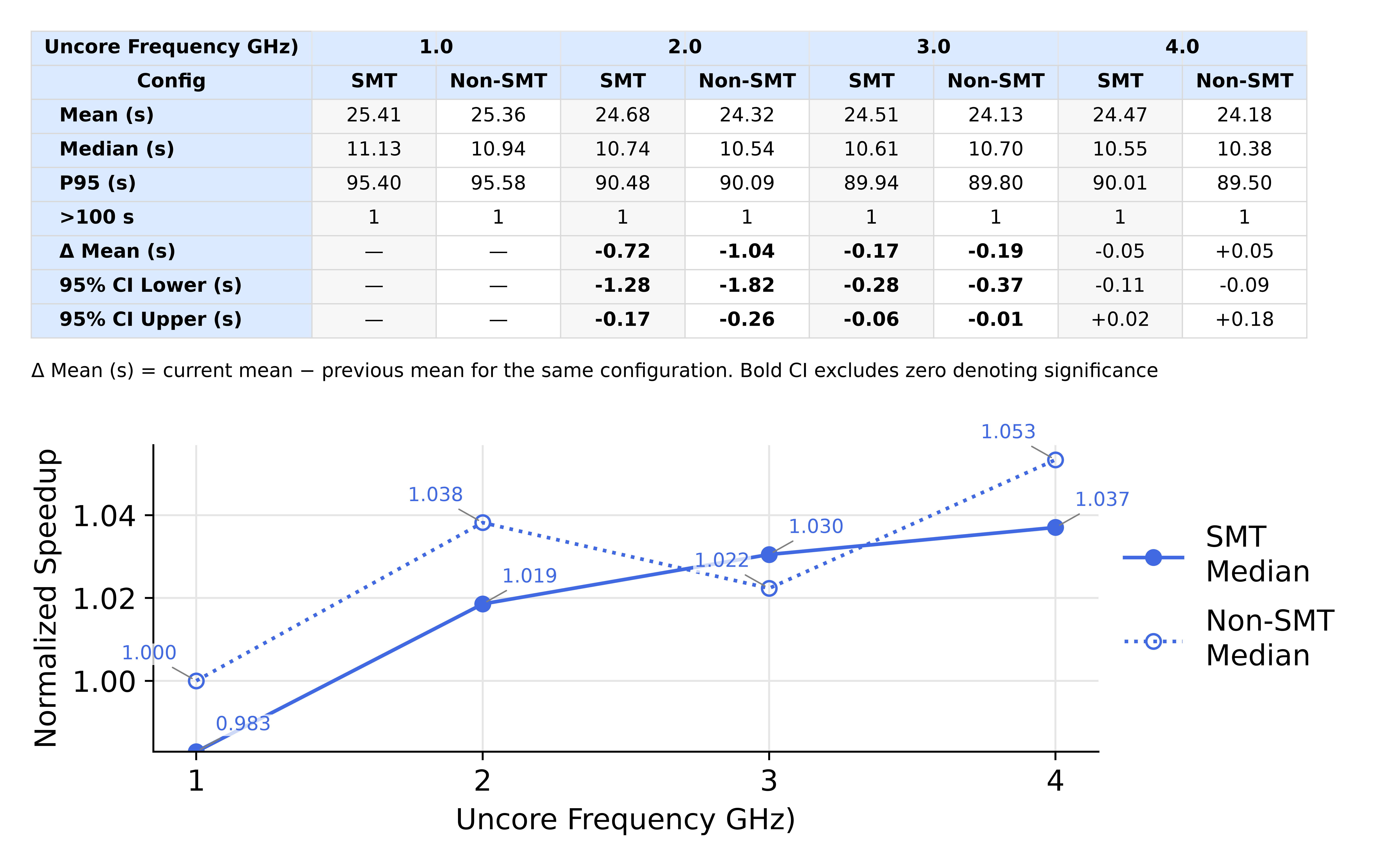}
    \vspace{-\figCaptionSpacing em}
    \caption{Single agent with increasing uncore frequency.}
    \vspace{-\figCaptionSpacing em}
    \label{fig:single_agent_uncore_freq_sweep}
\end{figure}

\textbf{Experiment 3: Single agent uncore frequency sensitivity.} 
This experiment evaluates four uncore frequencies using the \texttt{intel\_uncore\_frequency} driver: 1 GHz, 2 GHz, 3 GHz, and 4 GHz. This test is conducted on an eight P-core allocation with a core frequency of 4.5 GHz. 

To characterize the memory-system effects of uncore frequency, we measure idle memory-access latency using Intel Memory Latency Checker (MLC). When the core frequency is 4.5 GHz and uncore is 3 GHz (the frequency configuration for experiment 1), the average MLC idle memory-access latency for each logical P-core is 82.125 ns. At a core frequency of 4.5 GHz and an uncore frequency of 1 GHz, this average logical idle memory access latency increases to 116.85 ns. We then evaluate whether the workload is sensitive to increased memory path latency.

As shown in the speedup chart in Figure~\ref{fig:single_agent_uncore_freq_sweep}, increasing the uncore frequency from 1 GHz to 4 GHz with a single agent results in a 1.053$\times$ speedup for the non-SMT CPU allocation, while there is a dip in speedup at 3 GHz for the configuration without SMT utilization. Median speedup is normalized to the 1 GHz uncore configuration without SMT utilization. Taken together, there is a limited and inconsistent median task latency benefit for a single agent as uncore frequency increases from 1 GHz to 4 GHz. 

Configurations without SMT utilization exhibit slightly higher median speedups at 1, 2, and 4 GHz, while the SMT configuration performs better at 3 GHz. These differences do not exhibit a consistent relationship with uncore frequency. While there could be some sensitivity of SMT to uncore frequency, the effect appears to be minimal and inconsistent. 

\begin{figure}[!b]
\vspace{-5mm}
    \begin{takeaway}
        Increasing uncore frequency by 300\% produces at most a 1.053$\times$ median latency speedup for a single agent allocated eight P-cores, despite substantial changes in measured idle memory-access latency.
    \end{takeaway}
\vspace{-5mm}
\end{figure}

\newpage

\begin{figure}[t]
    \centering
    \includegraphics[width=1\linewidth]{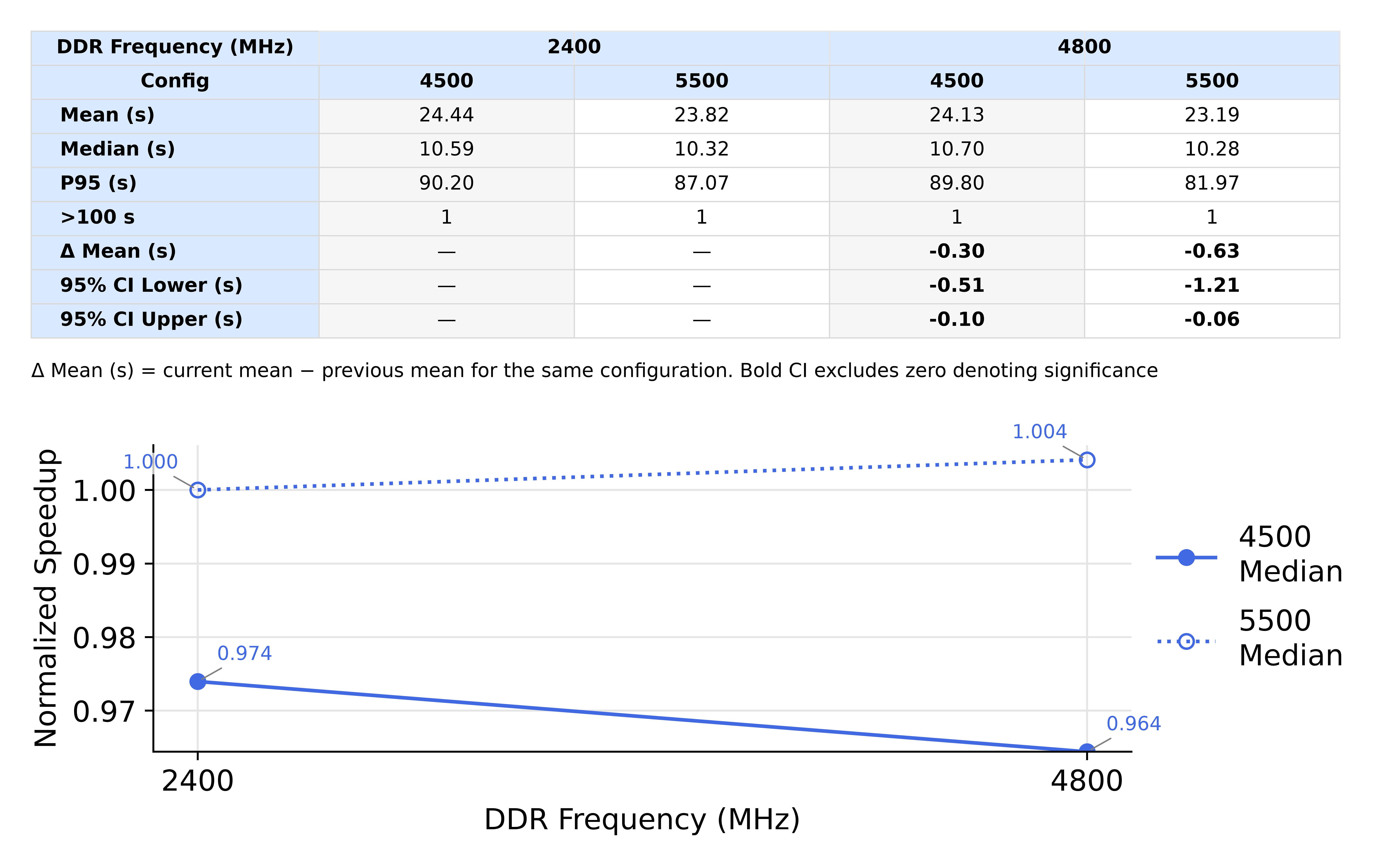}
    \vspace{-\figCaptionSpacing em}
    \caption{Single agent with different DRAM frequencies.}
    \vspace{-\figCaptionSpacing em}
    \label{fig:single_agent_dram}
\end{figure}

\textbf{Experiment 4: Single agent DRAM data-rate sensitivity.} 
Experiment 1 demonstrated that single agent CPU side task latency has minimal ($<1.05\times$) speedup with increased core counts. Experiment 3 demonstrated that single agent CPU side latency has minimal ($<1.05\times$) speedup improvement increasing uncore frequency from 1 GHz to 4 GHz. Only increasing core frequency by 120\% from 2.5 GHz to 5.5 GHz demonstrated a greater then 1.1$\times$ speedup in CPU side task latency. Together, these results motivate investigating whether the DRAM subsystem contributes to the limited performance scaling observed in single-agent execution.

We test whether the memory subsystem bandwidth and latency affect single agent performance on an eight P-core CPU allocation. For this test, we used the ASUS BIOS to reduce the memory frequency from 4800 MT/s to 2400 MT/s. Reducing the memory frequency increases the latency of retrieving data from memory, and reduces the bandwidth to main memory. 

Intel MLC tests show that reducing DRAM data rate from 4800 MT/s to 2400 MT/s reduces the average logical core idle latency from 116.85 ns to 107.8 ns with a core frequency of 4.5 GHz and an uncore frequency of 3 GHz.

The values in the speedup chart in Figure~\ref{fig:single_agent_dram} are all relative to the 2400 MT/s, 5.5 GHz median latency per task. At 5.5 GHz there is a minor speedup from 2400 MT/s to 4800 MT/s of 1.004$\times$. There is a slight slowdown from 2400 MT/s to 4800 MT/s of 0.974$\times$ to 0.964$\times$ at 4.5 GHz. The 4.5 GHz median task latency increased by 0.11 seconds, a 1.04\% increase, following a DRAM frequency change from 2400 MT/s to 4800 MT/s, a 100\% increase. These results suggest that single-agent task latency exhibits limited sensitivity to DRAM data rate across the evaluated configurations.

\begin{figure}[!b]
\vspace{-3mm}
    \begin{takeaway}
        Doubling DDR5 data rate from 2400 to 4800 MT/s changes median single-agent task latency by approximately 1\% or less across the evaluated P-core frequencies.
    \end{takeaway}
\vspace{-5mm}
\end{figure}

\newpage

\textbf{What affects single agent performance?} 

Through experiments 1-4 we have directly tested the major components of the computation and memory subsystems including:

\begin{itemize}
\vspace{-3mm}
    \item Physical CPU allocation
\vspace{-1mm}
    \item SMT logical cores
\vspace{-1mm}
    \item Physical core frequency
\vspace{-1mm}
    \item Uncore frequency
\vspace{-1mm}
    \item DRAM Frequency
\vspace{-3mm}
\end{itemize}

The only part of the subsystem which has an out sized affect ($>1.1\times $) on speedup is core frequency: an increase in core frequency from 2.5 GHz to 5.5 GHz imparts a 1.154$\times$ speedup. Increasing available physical cores for workload operation plateaus at 4 physical cores for a single agent, with a maximum speedup of 1.049$\times$. A single agent may be highly sequential and may not be constrained by limited parallel processing resources. Directly testing uncore and memory frequency demonstrates little effect; CPU side task latency of single agents may not be bandwidth or memory latency constrained. This raises the question: where in the computation and memory subsystems is the major bottleneck? 

The authors performed an additional 8 physical core, SMT off, 4.5 GHz core frequency, 3 GHz uncore, 4800 MT/s DRAM frequency test. We monitored this test with full VTUNE HPC profiling. The VTUNE profiling does not distinguish between sandbox setup time (requiring network access, memory capacity and bandwidth), and agentic task time. 

Our profiling results show that the workload may have been DRAM bandwidth bound for 6.1\% of workload clock ticks while the workload may have been cache bound for 18.1\% of clock ticks. Given we have established that there is little sensitivity of agentic task latency to memory frequency (which affects memory latency and bandwidth) and uncore frequency (which also affects latency and bandwidth to cores), these results indicate agentic workloads could be cache bound. VTune reports more cache-related stalls than DRAM-bandwidth-related stalls, suggesting that cache behavior warrants further investigation. While further investigation of the L1 and L2 cache is required, L1 and L2 cache bounding CPU side task latency aligns with the experiment 1 which found that core frequency can affect CPU side agentic task latency. The observed sensitivity to P-core frequency is consistent with execution phases whose latency depends on core-clock-domain operations, potentially including L1 and L2 cache accesses. However, additional profiling is required to distinguish cache-related stalls from other core-frequency-sensitive operations.

\newpage

\begin{figure}[t]
    \centering
    \includegraphics[width=1\linewidth]{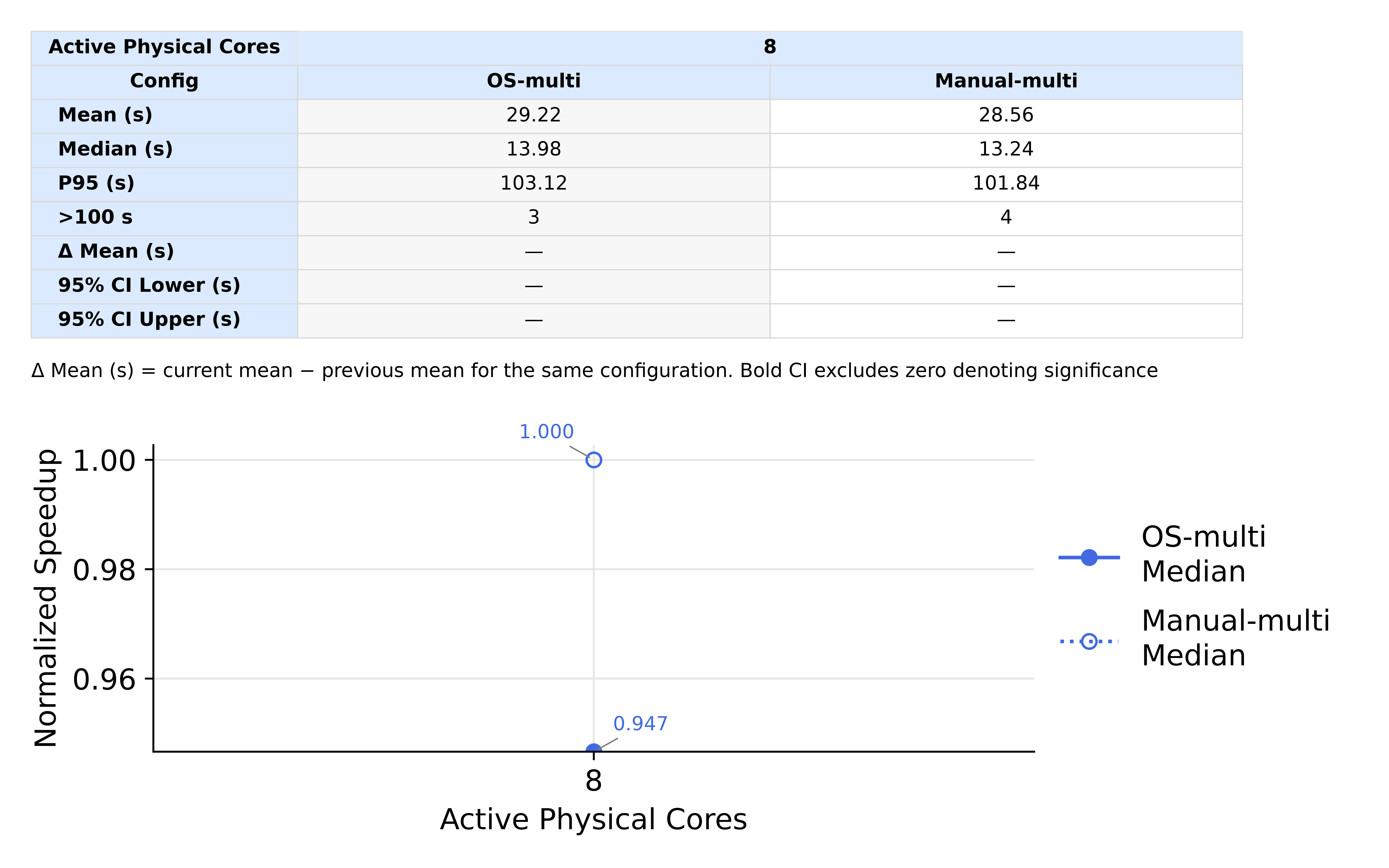}
    \vspace{-\figCaptionSpacing em}
    \caption{Multiple Agents with different software configurations.}
    \vspace{-\figCaptionSpacing em}
    \label{fig:multi_agent_os_configs}
\end{figure}

\textbf{Experiment 5: Comparing Multi-Agent Deployment Strategies.}
Our next series of experiments investigate CPU allocations running concurrent agents. For these multi-agent experiments, we allocate one agent for each active physical core, as shown with configuration two in Figure~\ref{fig:software_config}. If there are two physical cores active, we run two agent sessions concurrently. We compare two multi-agent deployment strategies to determine which minimizes CPU-side task latency on our system.

\begin{enumerate}
    \vspace{-4mm}
    \item \textbf{Partitioned:} Assign each agent session to a dedicated physical P-core using Docker CPU affinity and a separate Harbor instance.
    \vspace{-1mm}
    \item \textbf{Shared:} Launch one Harbor instance with concurrency equal to the allocated P-core count, allowing Linux to schedule agent sessions across the shared P-core pool.
    \vspace{-4mm}
\end{enumerate}

Chang et al.~\cite{chang2026llminferenceagenticworkloads} reports a 1.22$\times$ increase in mean execution time when concurrent agent sandboxes share a CPU allocation rather than use disjoint allocations. We investigate whether this pattern extends to our edge CPU and Harbor-based multi-agent deployment. 

To test manual assignment, we launch four Docker containers which launch individual Harbor sessions with a Docker-Out-Of-Docker (DooD) methodology. Each Harbor session completes one quarter of the total tasks. For our test, vLLM will run on all allocated resources.

Whereas Chang et al. allocate disjoint groups of logical CPUs to each sandbox, our partitioned configuration assigns one physical P-core per agent session. Both of the configurations tested in this experiment share the same L3 cache, so our comparison does not evaluate last-level cache isolation. We find that partitioned allocation has lower CPU side median and mean task latency; we apply this configuration in future multi-agent configurations.


\begin{figure}[!b]
\vspace{-4mm}
    \begin{takeaway}
        Consistent with~\cite{chang2026llminferenceagenticworkloads}, we find that partitioned CPU resources achieves lower task latency than shared allocation for concurrent agent execution.
    \end{takeaway}
\vspace{-4mm}
\end{figure}

\newpage

\begin{figure}[t]
    \centering
    \includegraphics[width=1\linewidth]{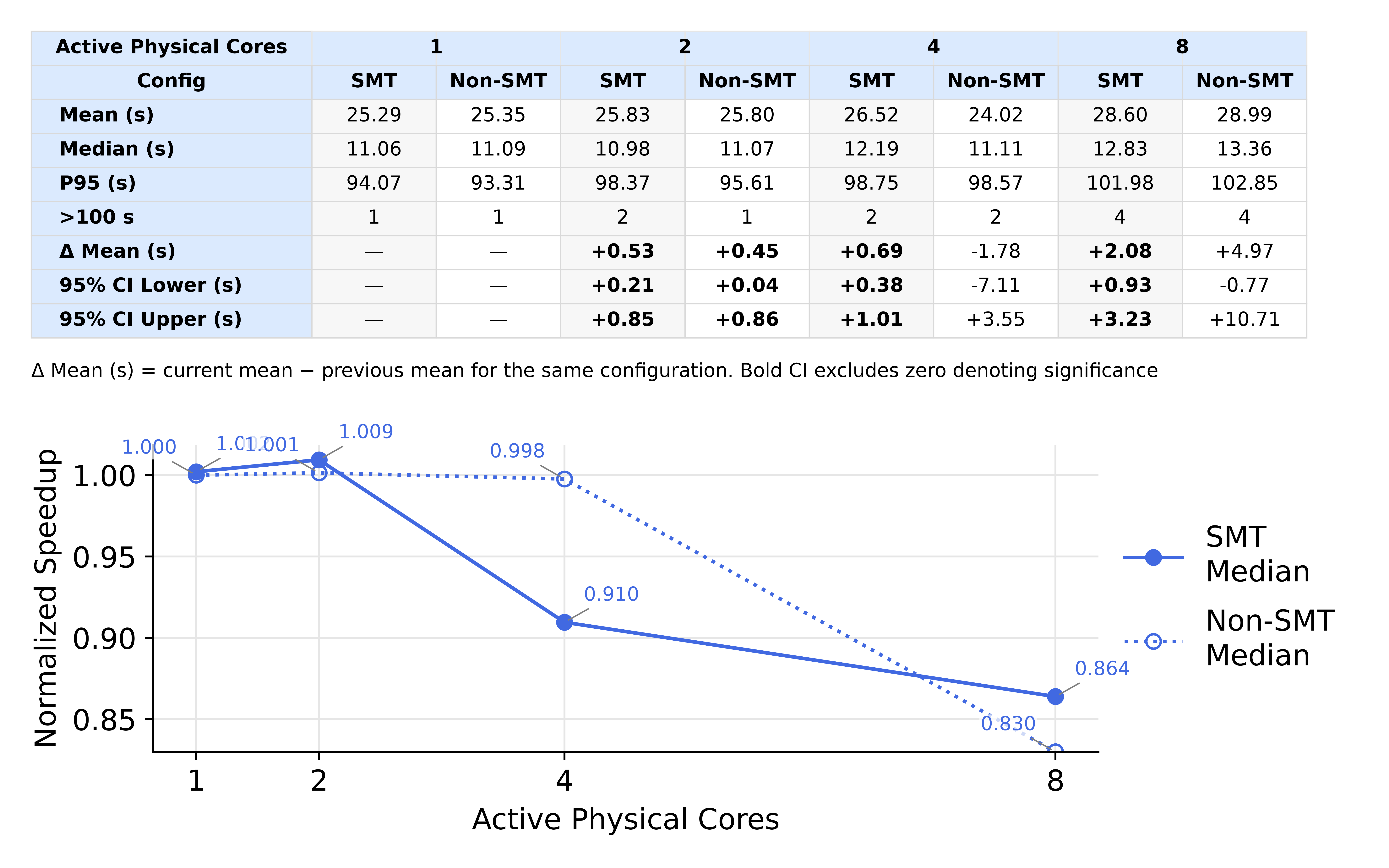}
    \vspace{-\figCaptionSpacing em}
    \caption{Multiple agents on increasing core allocations.}
    \vspace{-\figCaptionSpacing em}
    \label{fig:multi_agent_core_sweep}
\end{figure}

\textbf{Experiment 6: Multi-agent scaling with one P-core per agent.} Multiple agents running concurrently compete for shared CPU resources such as L3 cache, uncore bandwidth, and memory bandwidth. As the number of concurrent agents increases, contention for shared resources may increase, potentially affecting per-task latency even when each additional agent receives a dedicated P-core.

The upper table of Figure~\ref{fig:multi_agent_core_sweep} shows that mean and median task latency increase as the number of concurrent agent P-core pairs increases from two to eight. The 95th-percentile latency increases from 98.37 seconds to 101.98 for SMT configurations showing how additional concurrent agents impacts outlier tasks.

The lower plot of~\ref{fig:multi_agent_core_sweep} shows a steep slowdown in median task latency at 8 physical cores active relative to 4 cores for the non-SMT workload. Speedup drops from 0.998$\times$ of the non-SMT single core - single agent median task latency to 0.83$\times$ of the non-SMT single core - single agent median task latency. Beyond four P-core agent pairs, there is significant contention on the CPU causing performance degradation.

Interestingly, this multi-agent experiment is the first where SMT and non-SMT performance have a speedup difference greater than 0.05. With four physical cores with four agents running concurrently, the SMT median task latency increases to 12.19 seconds relative to 11.11 seconds for the non-SMT median task latency. At four concurrent agents, allocating both SMT siblings per P-core increases median task latency from 11.11 to 12.19 seconds, suggesting that exposing additional logical processors does not necessarily benefit partitioned multi-agent execution.

\begin{figure}[!b]
\vspace{-3mm}
    \begin{takeaway}
        Increasing concurrent agents proportionally with allocated P-cores does not preserve single-agent latency. At eight concurrent agents on eight P-cores, normalized median speedup decreases to 0.864$\times$ with SMT utilization and 0.830$\times$ without SMT utilization, indicating degraded per-agent performance despite dedicated physical core allocation.
    \end{takeaway}
\vspace{-2mm}
\end{figure}

\newpage

\begin{figure}[t]
    \centering
    \includegraphics[width=1\linewidth]{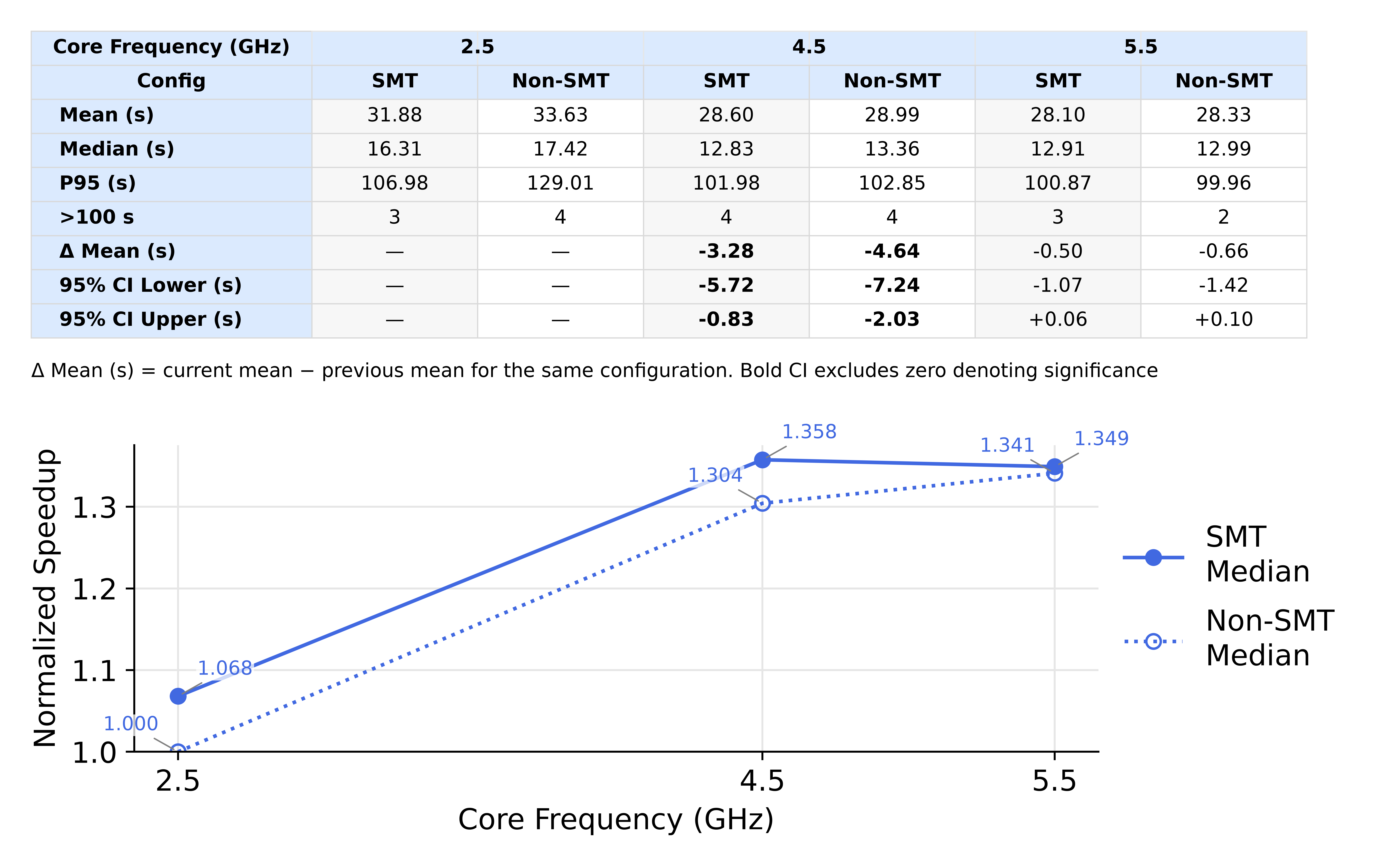}
    \vspace{-\figCaptionSpacing em}
    \caption{1:1 Core to agent with increasing Core frequency.}
    \vspace{-\figCaptionSpacing em}
    \label{fig:multi_agent_core_freq_sweep}
\end{figure}

\textbf{Experiment 7: Multi-agent P-core frequency sensitivity.} While experiments 1-4 characterized single-agent execution, we perform the next two experiments with eight concurrent agents on eight allocated P-cores. While core frequency had some impact and uncore frequency had little impact on single agents, concurrent agents present an opportunity to recover performance lost due to resource contention.

When all P-cores operate at 2.5 GHz, task latency increases substantially relative to the 4.5 GHz configuration. The upper table of Figure~\ref{fig:multi_agent_core_freq_sweep} contains 95th-percentile task latencies, showing longer task latencies for outlier tasks relative to the higher frequencies tested. The mean task latency for the non-SMT workload at 2.5 GHz is 33.63 seconds while the 4.5 GHz workload is 28.99, a 16\% increase. The median task latency increase from 4.5 GHz to 2.5 GHz is 25.9\% for the non-SMT configuration; the median increasing by a greater magnitude than the mean indicates that all task latencies are driving the slowdown, not just tail end tasks.

Median task latency speedup in Figure~\ref{fig:multi_agent_core_freq_sweep} decreases from 4.5 GHz to 5.5 GHz with SMT, and increases from 1.304$\times$ to 1.341$\times$ without SMT. The smaller improvement without SMT utilization and slight regression with SMT utilization between 4.5 and 5.5 GHz suggest diminishing latency benefits at higher P-core frequencies. The speedup caused by increased core frequency for multi-agent contrasts with speedup with a single agent. A single agent only has a 1.154$\times$ speedup when increasing to 5.5 GHz in non-SMT mode while increasing to 5.5 GHz with the multi-agent configuration has a 1.341$\times$ speedup. This indicates that multiple agents running concurrently are more sensitive to core frequency than a single agent.

\begin{figure}[!b]
\vspace{-5mm}
    \begin{takeaway}
        Increasing core frequency from 2.5 GHz to 5.5 GHz with 8 physical cores running 8 agents concurrently imparts a 1.341$\times$ speedup for the non-SMT configuration. Multi-agent task latency exhibits greater sensitivity to core frequency than single agent task latency.
    \end{takeaway}
\vspace{-2mm}
\end{figure}

\newpage

\begin{figure}[t]
    \centering
    \includegraphics[width=1\linewidth]{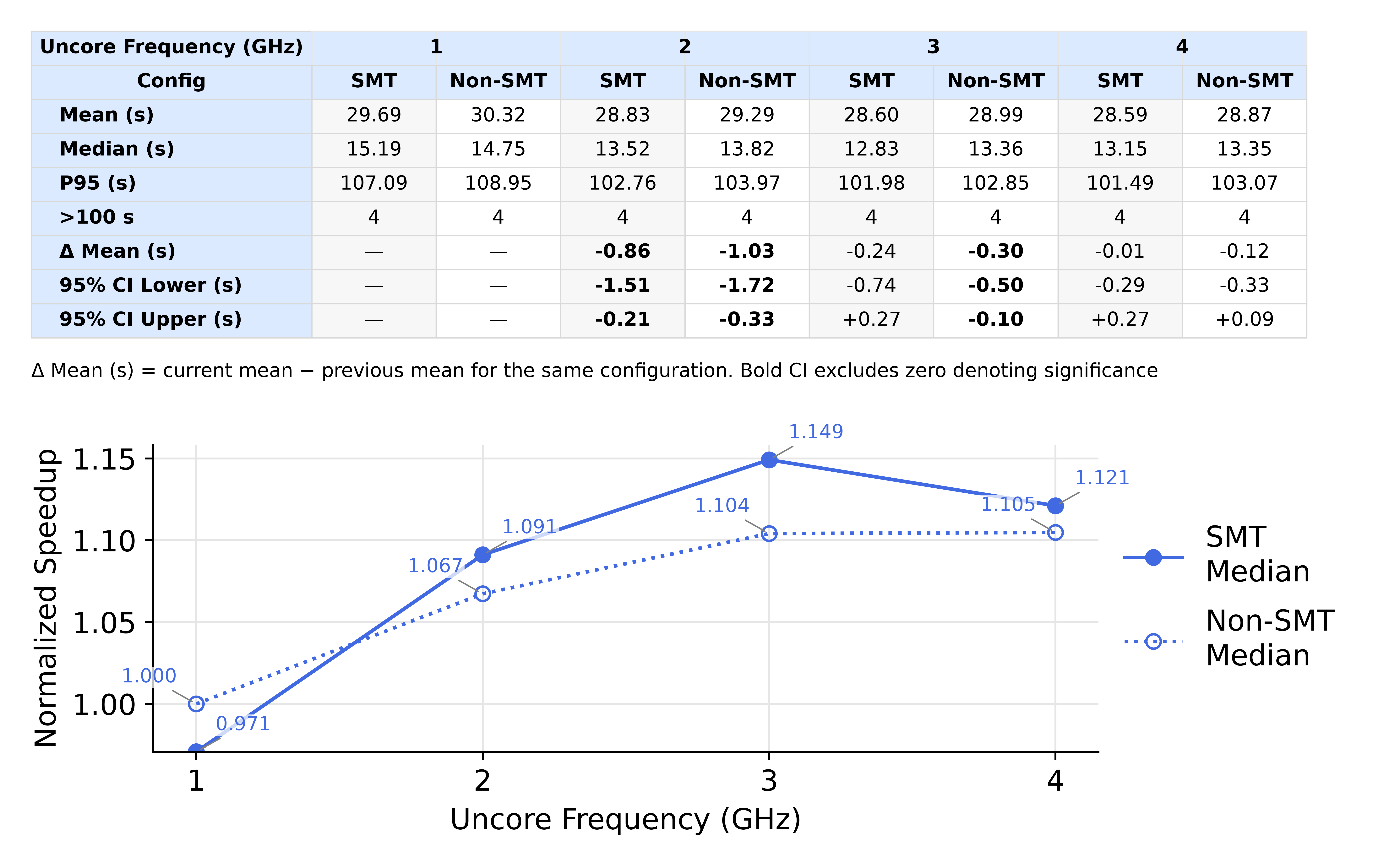}
    \vspace{-\figCaptionSpacing em}
    \caption{1:1 Core to Agent with increasing uncore frequency.}
    \vspace{-\figCaptionSpacing em}
    \label{fig:multi_agent_uncore_freq_sweep}
\end{figure}

\textbf{Experiment 8: Effects of changing uncore frequency with multiple agents.} With eight concurrent agents on eight P-cores running at 4.5 GHz, we test how changing the frequency of the uncore affects task latency. 

The speedup chart in Figure~\ref{fig:multi_agent_uncore_freq_sweep} shows a 1.149$\times$ median task latency speedup at 3 GHz uncore frequency with SMT relative to the 1 GHz non-SMT configuration. However, the speedup improvements plateau from 3 GHz to 4 GHz indicating minimal additional speedup gains from higher uncore frequency. 

The uncore driven speedup with concurrent agents stands in contrast to the single agent median task latency. Under concurrent execution, agentic task latency exhibits greater sensitivity to uncore frequency than under single-agent execution, suggesting that CPU operating-point requirements depend on agent concurrency. This motivates an uncore conscious design and configuration for concurrent agent CPU serving.

\begin{figure}[h]
\vspace{-3mm}
    \begin{takeaway}
        An eight P-core, 4.5 GHz SMT allocation with an uncore frequency of 3 GHz improves agentic task median latency by 1.149$\times$ relative to the non-SMT, 1 GHz baseline, motivating uncore analysis for concurrent agent serving.
    \end{takeaway}
\vspace{-8mm}
\end{figure}

\section{Discussion and Limitations} \label{sec:limitations}

We expose CPU side bottlenecks for single agent and multi agent operation, finding that single agent and multi agent CPU side operation is most sensitive to core frequency. Multi agent operation is more sensitive to core frequency than single agent operation. Agentic operation in the cloud uses larger core count server parts; the sibling to the Raptor Cove part we test is the Intel Emerald rapids server part which has a Max Turbo Frequency of 4.2 GHz for an 8 core count part, and a max turbo of 3.9 GHz for a 64 core part~\cite{EMR}. These findings motivate considering per-core frequency alongside total core count when provisioning CPUs for agentic workloads, particularly at low agent concurrency. High-frequency, lower-core-count processors may offer favorable per-task latency, although direct cross-platform measurements are required to establish this tradeoff.

Analyzing single agent task latency, we probed two points on the memory subsystem chain: DRAM frequency and uncore frequency. We measured the change in latency when changing both characteristics finding that our reduced DRAM frequency incurred roughly 25 ns of additional latency and our lowest uncore frequency imparted over 30 ns of additional latency. In either case, there are minimal effects of increased memory latency or reduced bandwidth on CPU side agentic task latency; exploratory VTune profiling identifies cache-related stalls as a potential contributor to CPU-side execution time. However, the measurements do not isolate the contribution of individual cache levels, motivating further micro architectural characterization.

Our work is based on a single LLM operating on the Terminal Bench benchmark conducted with the Harbor Framework. Total benchmarking time for our experiments is over four days; benchmark duration is a limitation exposed in other works which choose to reduce their samples from benchmarks to capture the benchmark in a timely manner~\cite{yuan2026agenticaiworkloadcharacteristics}. Due to this limitation, we only characterize CPU performance on coding agentic tasks with Terminal Bench. While much of the software stack is shared between agentic flows, such as tokenization, orchestration, and agentic harness, the agent tools calls and behavior can differ, motivating further CPU characterization for other classes of agentic workloads.

Our trajectory-replay methodology improves reproducibility by eliminating variation in LLM-generated actions and controlling synthetic inference delays. However, replay circumvents LLM inference, excluding GPU execution and associated CPU side LLM work. Furthermore, our synthetic inference latency model approximates a B200 based on a Hugging Face study of a similar LLM to our Qwen3~\cite{jin2026qwen3omni}. Consequently, our findings may differ in live inference deployments.

\section{Conclusion} \label{sec:conclusion}

We perform 8 experiments to characterize the effects of micro-architectural changes on CPU side latency for agentic tasks. We analyze bottlenecks for both single agent operation and when multiple agents run concurrently. We find core frequency to be the most effective mechanism to change agentic task latency while memory latency changes have little affect on performance. These findings warrant further micro-architectural investigation into other agentic tasks such as chat, web search, and data processing. We produce an open source dataset and framework for replayable agentic benchmarks to enable future CPU side investigation.


\bibliographystyle{plainnat}
\bibliography{citations}

\end{document}